# Fine-scale observations of the Doppler frequency shifts affecting meteor head radio echoes

Jean-Louis Rault[1], Mirel Birlan[2], Cyril Blanpain[3], Sylvain Bouley[4], Stéphane Caminade[4], François Colas[2], Jérôme Gattacceca[5], Simon Jeanne[2], Julien Lecubin[3], Adrien Malgoyre[3], Chiara Marmo[4], Jérémie Vaubaillon[2], Pierre Vernazza[6], Brigitte Zanda[7]

[1] International Meteor Organization (IMO) radio commission, 91360 Épinay-sur-Orge, France
f6agr@orange.fr

[2] Institut de Mécanique Céleste et de Calcul des Éphémérides (IMCCE), Observatoire de Paris, 75014 Paris, France
mirel.birlan@obspm.fr, francois.colas@obspm.fr, simon.jeanne@obspm.fr, jeremie.vaubaillon@obspm.fr,

[3] OSU Pythéas, CNRS, Université d'Aix-Marseille, 13007 Marseille, France
blanpain@osupytheas.fr, sylvain.bouley@gmail.com, julien.lecubin@osupytheas.fr, adrien.malgoyre@osupytheas.fr,

[4] Géosciences Paris Sud (GEOPS), Université Paris Sud, CNRS, Université Paris-Saclay, 91405 Orsay, France
stephane.caminade@ias.u-psud.fr, chiara.marmo@u-psud.fr,

[5] Centre Européen de Recherche et d'Enseignement des Géosciences de l'Environnement (CEREGE), CNRS, Aix-en-Provence, France
gattacceca@cerege.fr

[6] Laboratoire d'Astrophysique de Marseille, Université d'Aix-Marseille, 13007 Marseille, France
pierre.vernazza@lam.fr

[7] Muséum National d'Histoire Naturelle, 75005 Paris, France
brigitte.zanda@mnhn.fr

The French FRIPON (Fireball Recovery and Interplanetary Observation Network) programme relies on a video cameras network associated to radio sensors running in a radar multistatic configuration to observe fireballs and to determine meteoroid accurate orbits and potential meteorites strewnfields. This paper focuses on some peculiar phenomena observed with radio means during the final phase of the meteors flight.

## 1  Introduction

The French FRIPON (Fireball Recovery and Interplanetary Observation Network) programme plans to install a 100 video cameras and 25 radio receivers network to observe fireballs in order to compute associated meteoroid orbits and to determine the eventual meteorites strewnfield. Currently, 80 cameras and 13 receivers are already operational.

Meteoroid orbit and atmospheric trajectory is determined by means of optical triangulations on the bolide apparent trajectories, and its accurate velocity is computed thanks to radio data.

The French Air Force GRAVES radar, which primary purpose is to detect and classify satellites, is used by FRIPON in a multistatic configuration running in forward and back scatter modes.

Although the main initial purpose of the radio observation system was the calculation of the accurate velocity of the meteoroids, it quickly became apparent that this system also enabled, as a result of its properties, the detailed observation of the meteoroids behavior during their atmospheric flight.

## 2  Observational method

A multistatic radar configuration has been chosen for the FRIPON experiment.

The GRAVES transmitter, located near Dijon, France is a HPLA (High Power Large Aperture) radar-type. It is transmitting 24 hours a day a powerful permanent 143.050 MHz CW (Continuous Wave) carrier thanks to its 4-planar phased-array antennas. Each of the 4 antenna systems is scanning an azimuthal sector of 45 ° (from 90° to 270°) in the southern direction of Dijon.

The FRIPON SDR (Software Defined Radios) receivers are located where some of the 100 video cameras are (see figure 1), and they share the same local computer that is used to process and to transfer the video data to the FRIPON central server and database. Each radio set-up consists in a FunCube Pro +2 SDR receiver and a colinear vertical omnidirectional antenna.

Each time an optical multi-detection occurs (i.e. a meteor is detected on several video camera stations), the related video and SDR I/Q (In phase / in Quadrature) radio data is automatically transferred through a VPN (Virtual Private Network) to the FRIPON central server. Furthermore, the I/Q radio data recorded 24 hours a day are stored for about one month on the local hard drive of each station and can thus be uploaded at any time for data processing, even if no optical event is associated.

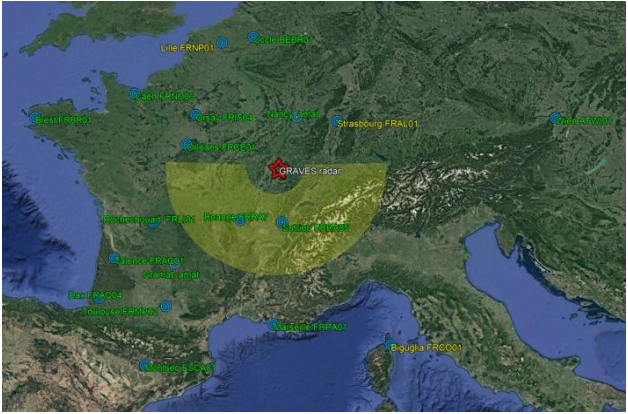

*Figure 1* – Location of the GRAVES transmitter FRIPON the radio receivers. In addition to the French receivers, some receivers are also installed in Austria, Spain and Belgium (Sept. 2017).

## 3   Head echoes observations

The radio waves radiated by a transmitter are scattered by the free electrons of the plasma surrounding the fast moving body of a meteoroid and by the free electrons of its ionized train, provided that the density of these electrons is high enough, depending on the frequency of the radio waves that is used. The plasma frequency below which a radio wave is reflected is given by:

$$f_n = \sqrt{\frac{N_e \cdot e^2}{\Pi \cdot m}} \qquad (1)$$

$m$ and $e$ being respectively the mass and the electric charge of an electron, $N_e$ the electron density and $f_n$ the frequency of the reflected wave.

A trail echo amplitude is very sensitive to the geometrical configuration of the incident radio waves. The Snell Descartes law fully applies to the quasi cylindrical train mirror that produces a specular reflection. If a meteor train is not properly oriented (i.e. if it is not tangent to an ellipse which foci are the TX and RX locations), there will be no train echo at all.

On the other hand, a head echo has much less geometrical dependence, as a large part of the reflecting plasma surface surrounding the meteoroid is considered as almost spherical, as observed by Kero et al. (2008) and simulated by Dyrud et al. (2008). Therefore the RCS (Radar Cross Section) of a head echo look similar for different receivers located at distant places.

## Usual head echoes used for meteoroid velocity measurement

Knowing at any time the position of a meteoroid entering the atmosphere thanks to its trajectory determination with video means, its velocity is deducted from (2) or (3).

$$f_{RX} = f_{TX} \frac{1 - \frac{dR_T}{cdt}}{1 + \frac{dR_R}{cdt}} \qquad (2)$$

$$f_{RX} = f_{TX} \frac{1 + \frac{V_g}{c} \cdot \cos(\delta - \frac{\beta}{2})}{1 - \frac{V_g}{c} \cdot \cos(\delta + \frac{\beta}{2})} \qquad (3)$$

where $V_g$ is the geocentric velocity of the target, *TX* the transmitter location, *RX* a receiver location, $f_{TX}$ the transmitted frequency and $f_{RX}$ the Doppler shifted frequency measured at the receiver location.

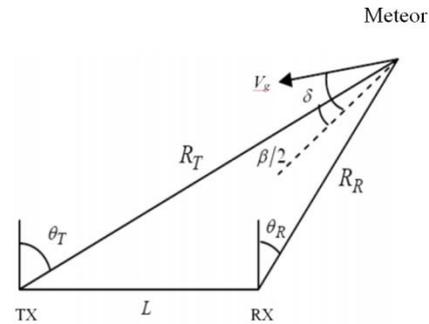

*Figure 2* – TX: transmitter location, RX: receiver location, $V_g$: Geocentric target velocity vector

Figure 3 shows an example of such a meteor head echo used to compute its geocentric velocity.

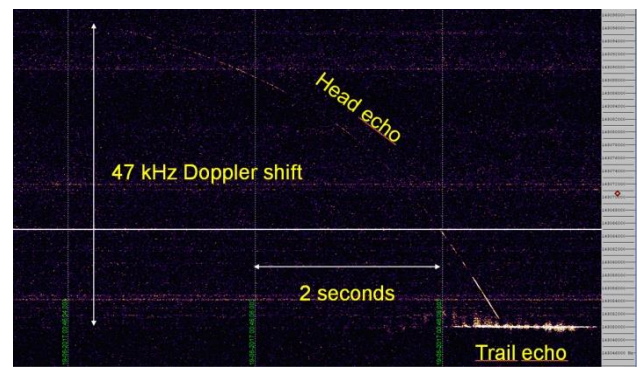

*Figure 3* – Example of fireball head echo detected on 19 june 2017 at $00^h\ 46^m\ 08^s$ UTC on the FRIPON radio station in Toulouse.

We find that on a large number of head echoes, the Doppler frequency drift consists in a smooth curve. The trajectory of the meteoroid shown on figure 3, computed thanks to the video data from 5 FRIPON cameras, is displayed on figure 4. Figure 5 shows the related measured magnitudes of the meteor along its path.

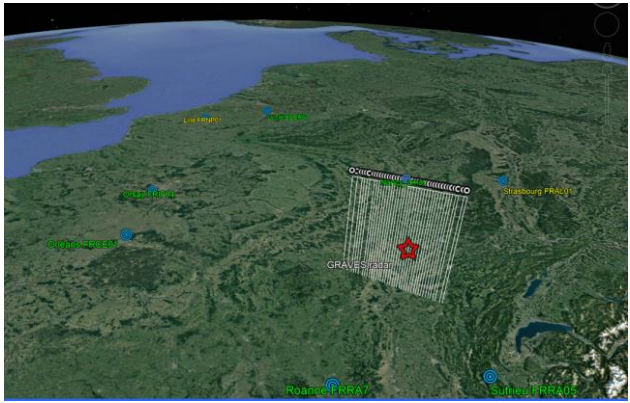

*Figure 4* – Trajectory of the 19 June 2017, $00^h\ 46^m\ 08^s$ UTC bolide computed from video data of FRIPON camera stations located in Besançon, Chatillon, Pontarlier, Saint Lupicin and Troyes.

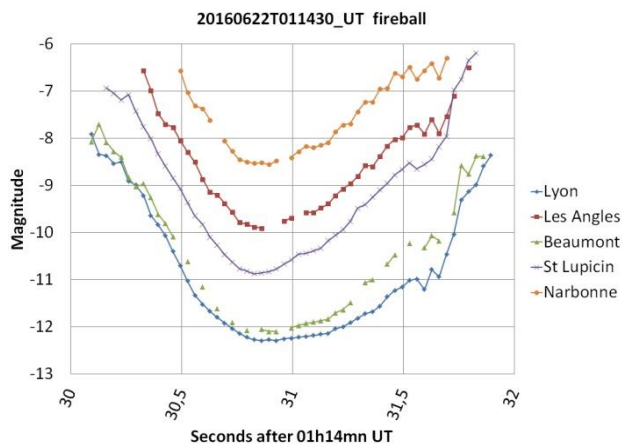

*Figure 5* – Magnitudes graphics of the 19 June 2017 bolide as recorded by 5 different FRIPON cameras.

The main purpose of FRIPON radio observations is to determine as accurately as possible the meteoroids velocity. However, it appeared serendipitously that the HPLA configuration of the transmitter used by FRIPON, associated with the facts that it is transmitting a pure CW VHF (Very High Frequency) carrier allow to observe a variety of detailed meteor heads phenomena, such as fragmentation, spinning meteoroid, and others.

**Example of a partial fragmentation of a bolide**
Sometimes a meteoroid fragments partially during its atmospheric trajectory. An example of such a behavior can be seen on figure 6, for a fireball recorded on 22 June 2016 at $01^h\ 14^m\ 30^s$ UTC.

.

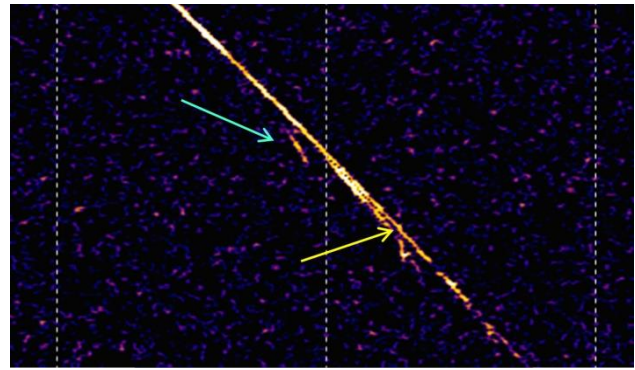

*Figure 6* – The arrows show at least two distinct fragments decelerating faster than the main body observed on 22 June 2016 at $01^h\ 14^m\ 30^s$ UTC.

**Sudden changes of head echoes Doppler shifts**
Smooth Doppler shifts curves indicate that the radial velocity measured by each radio station is varying regularly, according to the geometry of figure 2. But it happens that large and sudden variations occur during the final phase of a fireball life.

A first example of such a behavior is shown on figure 7

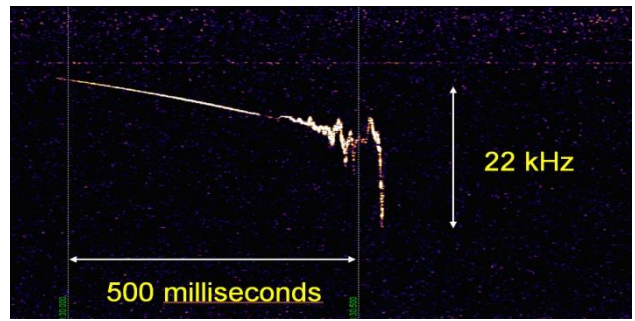

*Figure 7* – Sudden Doppler shift variations on the bolide observed on 15 July 2017 at $02^h\ 01^m\ 30^s$ UTC

**Regular oscillations on head echoes Doppler shifts**
It appears that meteor Doppler shift curves are sometimes affected by regular oscillations during the last phase of their trajectory, as seen on a first example on figures 8 and 9.

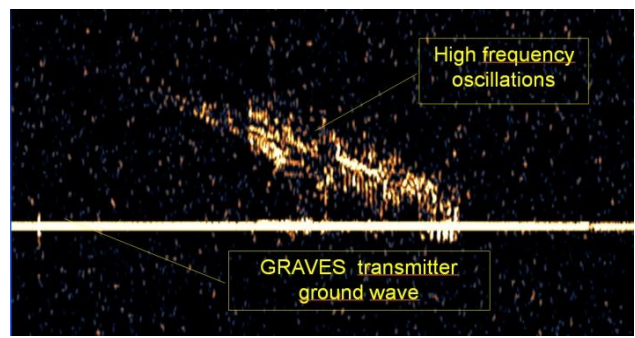

*Figure 8* – High frequency oscillations affecting the Doppler shift curves of the bolide detected on 4 August 2017 at 00h 04m 46s UTC, as recorded by the FRIPON radio station located in Suttrieu. The same phenomenon was also observed by the distant radio station of Toulouse.

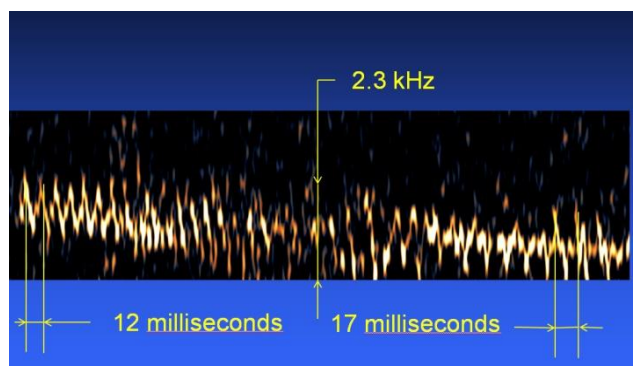

*Figure 9* – Zoom on the Doppler shift oscillations shown on the previous figure 8.

Finally, a complex Doppler signature of a fireball observed on 16 December 2016 at $01^h 32^m 21^s$ UTC is shown on figure 10.

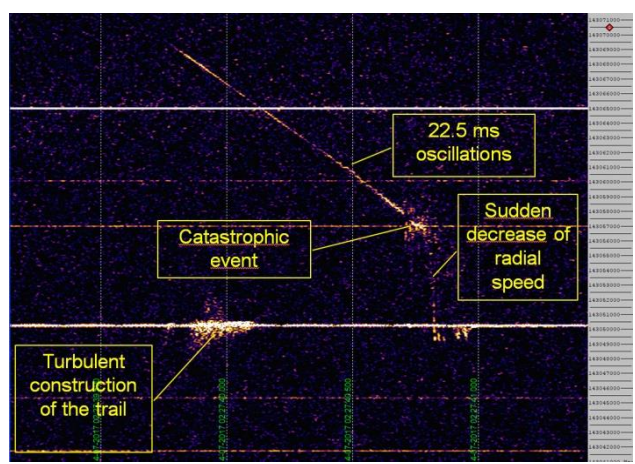

*Figure 10* – Doppler signature of the bolide observed on the $16^{th}$ of December 2016 at $01^h 32^m 21^s$ UTC by the Orléans, Marseilles and Orsay FRIPON radio stations

Such high frequency periodic fluctuations of the Doppler frequency shifts are not observed in any case by the FRIPON cameras. The fact that their shooting rate is only 30 frames per second probably explains that the cameras do not see such fast magnitude variations.

## 4 Discussion

Meteor head plasma behavior simulations such as performed by Dyrud et al. (2008) and Silber et al. (2017) apply to meteor flying in a steady state. At the end of their flights, bolides undergo very harsh stresses due to the exponential increase of the atmospheric pressure, stresses which deeply modify the behavior of the plasma surrounding the meteoroid. The sudden frequency variations in Doppler signatures observed with the FRIPON radio system, that employs a HPLA radar in multistatic mode, highlights large radial velocities and associated RCS fluctuations. In the case for example of figure 10, it appears that the sudden frequency variations are due to some changes in the apparent distance of the RCS, probably because the global volume and the shape of plasma surrounding the body changes during the final fragmentation.

The regular Doppler oscillations visible on figures 8 and 9 can be explained, according to the current observational data, by the spinning of a dissymmetric meteoroid. This hypothesis is supported by some meteor magnitude flickering already observed by Beech et al. (2003), and by the periodic RCS amplitude fluctuations detected by Kero et al. (2005).

## 5 Conclusion

A HPLA transmitter such as GRAVES, transmitting a permanent pure carrier at a short wave length (about 2 m) proves to be a powerful tool to examine in detail the plasma surrounding the meteoroid. The automation of our data processing of radio records will allow the systematic analysis of all recorded head echoes. The combination of radio data from different stations allowing a 3D vs time view of the observed phenomena are some of the next steps ahead for the FRIPON programme that is planned to run for the next 10 years.

## Acknowledgment

FRIPON is funded by the ANR grant N.13-BS05-0009-03. FRIPON data are hosted at IDOV+C (Integrated Data and Operation Center). Support for IDOC is provided by CNRS and CNES. Warm thanks to Karl Antier for the attentive proofreading of this paper and for his pertinent remarks.